\documentclass[draftcls,12pt,onecolumn,oneside]{IEEEtran}
\usepackage{mathrsfs}
\usepackage{algorithm}
\usepackage{algorithmic}
\usepackage{url,times,amsmath,color,amssymb,graphicx,epsfig,psfig,cite,geometry,psfrag,subfigure,dsfont,booktabs, multirow}
\begin{document}
\title{Self-Organization in Disaster Resilient Heterogeneous Small Cell Networks}
\author{Haijun Zhang,~\IEEEmembership{Member,~IEEE}, Chunxiao Jiang,~\IEEEmembership{Member,~IEEE}, Rose Qingyang Hu,~\IEEEmembership{Senior Member,~IEEE}, and Yi Qian,~\IEEEmembership{Senior Member,~IEEE}
\thanks{Haijun Zhang is with College of Information Science and Technology, Beijing University of Chemical Technology, Beijing 100029, P. R. China, and is also with the Department of Electrical and Computer Engineering, the University of British Columbia, Vancouver, BC, V6T 1Z4, Canada  (Email: dr.haijun.zhang@ieee.org).

Chunxiao Jiang is with the Department of Electronic Engineering, Tsinghua University, Beijing
100084, P. R. China (Email: jchx@tsinghua.edu.cn).

Rose Qiangyang Hu is with Department of Electrical and Computer Engineering, Utah State University, Logan, UT, USA (Email: rosehu@ieee.org).

Yi Qian is with the Department of Electrical and Computer Engineering, University of Nebraska-Lincoln, Omaha, NE,
USA (e-mail: yi.qian@unl.edu, Corresponding Author).

}}
\maketitle

\begin{abstract}
Heterogeneous small cell networks with overlay femtocells and macrocell is a promising solution for future heterogeneous wireless cellular communications. However, great resilience is needed in heterogeneous small cells in case of accidents, attacks and natural
disasters. In this article, we first describe the network architecture of disaster resilient heterogeneous small cell networks (DRHSCNs), where several self-organization inspired approaches are applied. Based on the proposed resilient heterogeneous small cell network architecture, self-configuring (power, physical cell ID and neighbor cell list self-configuration) and self-optimizing (coverage and capacity optimization and mobility robustness optimization) techniques are investigated in the DRHSCN. Simulation results show that self-configuration and self-optimization can effectively improve the performance of the deployment and operation of the small cell networks in disaster scenarios.
\end{abstract}

\section{Introduction}
Wireless communications networking continues to play an increasingly important role in our daily life. A small cell base station, e.g., a femtocell, which is also known as a HeNB, is a low power wireless access point that can improve the capacity and coverage of indoor and hotspot scenarios \cite{RoseHuYiQianHetNet13}. Currently, heterogeneous small cell network with overlay femtocells and
macrocell is a most promising solution for the wireless cellular communications of the
future \cite{FemtoSpringer2013}.  However, accidents, attacks and natural disasters like earthquakes,  floods  and  hurricanes  are always occurring in many  places  around  the  world \cite{AdachiDisasterWPMC2012, ZhuHaojinTPDS2014}.  Therefore, disaster resilient heterogeneous small cell networks (DRHSCNs) need to be studied comprehensively.

Disaster resilient communication networks benefit from employing self-organization
networks (SON) greatly \cite{survivabilityIEEESurvey09}. The principal objective of introducing self-organization which comprises self-configuration, self-optimization, and self-healing, is to effect substantial operational expenditure (OPEX) reductions by reducing human involvement in network optimization, while improving network efficiency and quality of service (QoS). Considering the requirements of minimal human involvement in the network deployment and optimization tasks in disaster resilience communications, the need for SON enabled DRHSCNs is crucial.

Many researches on SON has been done recently. In \cite{SONIEEESurvey13},  the authors survey the literature
over the last decade on the emerging field of self organization as applied to
wireless cellular communication networks. In \cite{HonglinHuSONComMagazine2010}, the authors focus on the self-configuration and self-booting mechanism for a newly evolved NodeB (eNB), and a distributed mobility load balancing algorithm with low handover cost is proposed and evaluated. In \cite{MugenPengSONIEEEComMag13}, the state-of-the-art research on self-configuration and self-optimization in LTE-Advanced heterogeneous networks are surveyed. Self-optimization is investigated in LTE, femtocell, D2D and 5G network \cite{RoseHuYiQianLTE14, HaijunTCOM2014, RoseHuYiQianD2D14, RoseHuYiQian5G14}. However, self-organizing disaster communication networks have not been considered in the previous works.

SON in DRHSCNs consists of self-configuration, self-optimization and self-healing \cite{SONIEEESurvey13}. Additional base stations are self-configured in a `plug-and-play' fashion, while existing base stations continuously self-optimize their operational algorithms and parameters in response to changes in disaster communication networks with traffic and environmental conditions. The optimizations aim to provide the reliability and high QoS as efficiently as possible. In case of failure of  a cell or site, self-healing is triggered to resolve the resulting problem in coverage and capacity. Due to space limitation, we mainly focus on self-configuration, self-optimization in this paper.

The need for self-organization in DRHSCNs is driven by a number of technology challenges. The complexity is high for wireless radio access technologies, primarily due to the large number of tunable parameters and the dependencies among them, especially in disaster resilient heterogeneous small cell networks. This complexity also makes network operations more challenging. It is even more challenging with the large number of sites/cells required to provide coverage with future high-frequency technologies, and the coexistence and coordinated exploitation of multiple heterogeneous access networks.

Due to these technological complexities, the key operational tasks of radio network deployment, configuration,  planning and optimization are severely unsatisfied in disaster scenarios. Intrinsic problems and shortcomings of the methodologies applied in heterogeneous small cells of disaster scenarios include: 1) Tremendous increase of traffic (e.g. voice calls)  versus  the limited capacity of the damaged wireless networks; 2) Damaged base stations with disabled backhaul; 3) Confusion and the conflict of the physical cell identity (PCI) for newly added base-stations; 4) Neighbor cell list configuration; 5) Handover related radio link failure (RLF) and unnecessary handovers; 6) Limited human resources in disaster rescue scenario versus the requirement of massive human intervention in disaster wireless network deployment and optimization.

In essence, the current approaches are characterized by a high labor-intensity, and these approaches only deliver largely suboptimal solutions in disaster wireless communication networks.  This is a key enabler to support self-organization objectives in disaster wireless communication networks. The key gains from employing self-organization in disaster small cell networks are in the forms of OPEX reductions and performance enhancements.
In the areas of drive testing, network planning, monitoring and optimization, human involvement is expected to reduce. Drive tests are currently performed to check the performance of the network. Measurements by user equipment (UE) and eNBs can reduce those labor-intensive tasks. Using measurements from UEs also have the advantage that measurements are obtained from additional locations, such as inside buildings.
Self-organization features also enhance network performance and service quality by better adapting to specific characteristics and requirements in disaster scenarios.

In this article, we apply self-organization approaches in disaster resilient heterogeneous small cells. A SON enabled DRHSCN architecture is introduced in Section II. Self-configuration, including PCI self-configuration and automatic neighbor relation (ANR) self-configuration, is studied in Section III. Self-optimization, including mobility robustness optimization and coverage and capacity optimization, is presented in Section IV. The conclusion of this article is in Section V.

\section{A SON enabled DRHSCN architecture}

\begin{figure}[h]
        \centering
        \includegraphics*[width=12cm]{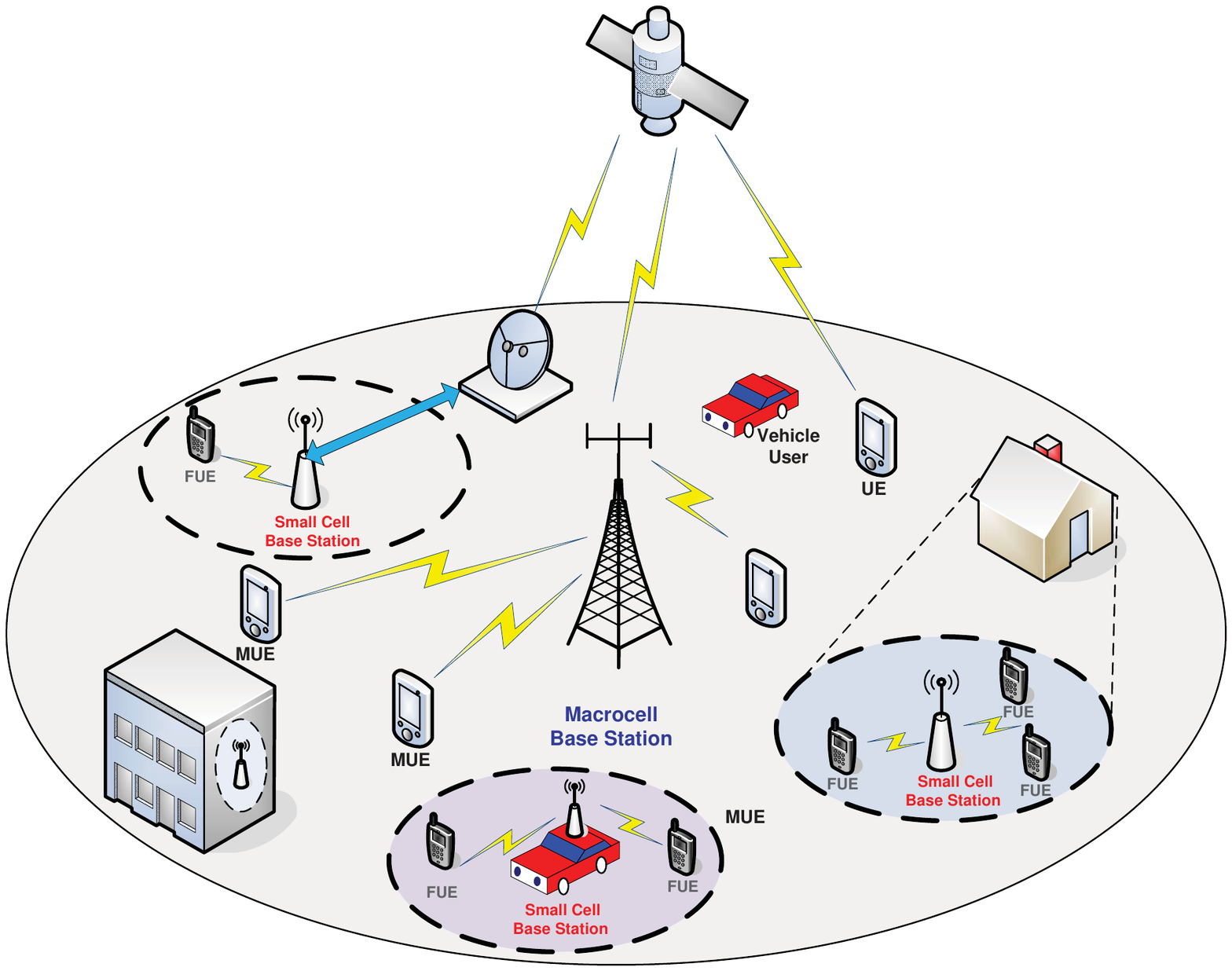}
        \caption{DRHSCN Network Architecture.}
        \label{fig:1}
\end{figure}

Fig. 1 shows the network architecture of a DRHSCN, which is comprised of a macrocell and several small cells, with the satellite access link as the backhaul. The SON entity can be deployed on the base station side. In this article, we focus on self-configuration and self-optimization in SON enabled  DRHSCNs.

A self-configuration process is defined as a process in which newly deployed nodes
in a disaster scenario are configured by automatic installation procedures to get the
necessary basic configuration for system operation \cite{3GPP36902}.
The self-configuration phase is triggered by `incidental events' or `intentional events'. Examples are the addition of a new site and the introduction of a service or a new network feature. In this case, it usually require eNB to configure many radio resource management parameters and schemes, neighbor lists and pilot powers. The process of configuration should be prior to the self-optimization.

A self-optimization process is defined as a process where UE and eNB measurements and performance optimization is self-optimized \cite{3GPP36902}.
Self-optimization can self-optimize the radio resource management parameters, including power settings (pilot, control and traffic channels), antenna parameters (tilt, azimuth), neighbor lists (PCI and associated weights), and a range of radio parameters (admission/handover management and scheduling). The process of self-optimization can work when the RF interface is switched on.

The degree of self-organization in cellular network can be determined by network operators case by case. Generally, operator only need to determine the SON methods with certain guidance, and it usually requires balance with tradeoffs in coverage, capacity, QoS and cost. SON can provide operator with new deployed eNB information, advice of good deployment location, suggestions for new channel boards, type of amplifiers, or setting of antenna tilt angle, etc. Consider a fully configured and operating radio access network, which is arbitrarily start at the certain ``measurement" phase. In this phase, massive measurement of various parameters are collected by the eNB or UE's feed back. These collected measurement parameters of radio channel characteristics, traffic and users' mobility states, are processed before provided to SON entity for various  self-optimization tasks. The specific parameters, report period, etc., usually depend on the certain scheme of the self-optimization method.

\section{Self-Configuration in DRHSCNs}

\subsection{Self-Configuration of Transmitter Power}
\begin{figure}[h!]
        \centering
        \includegraphics*[width=16cm]{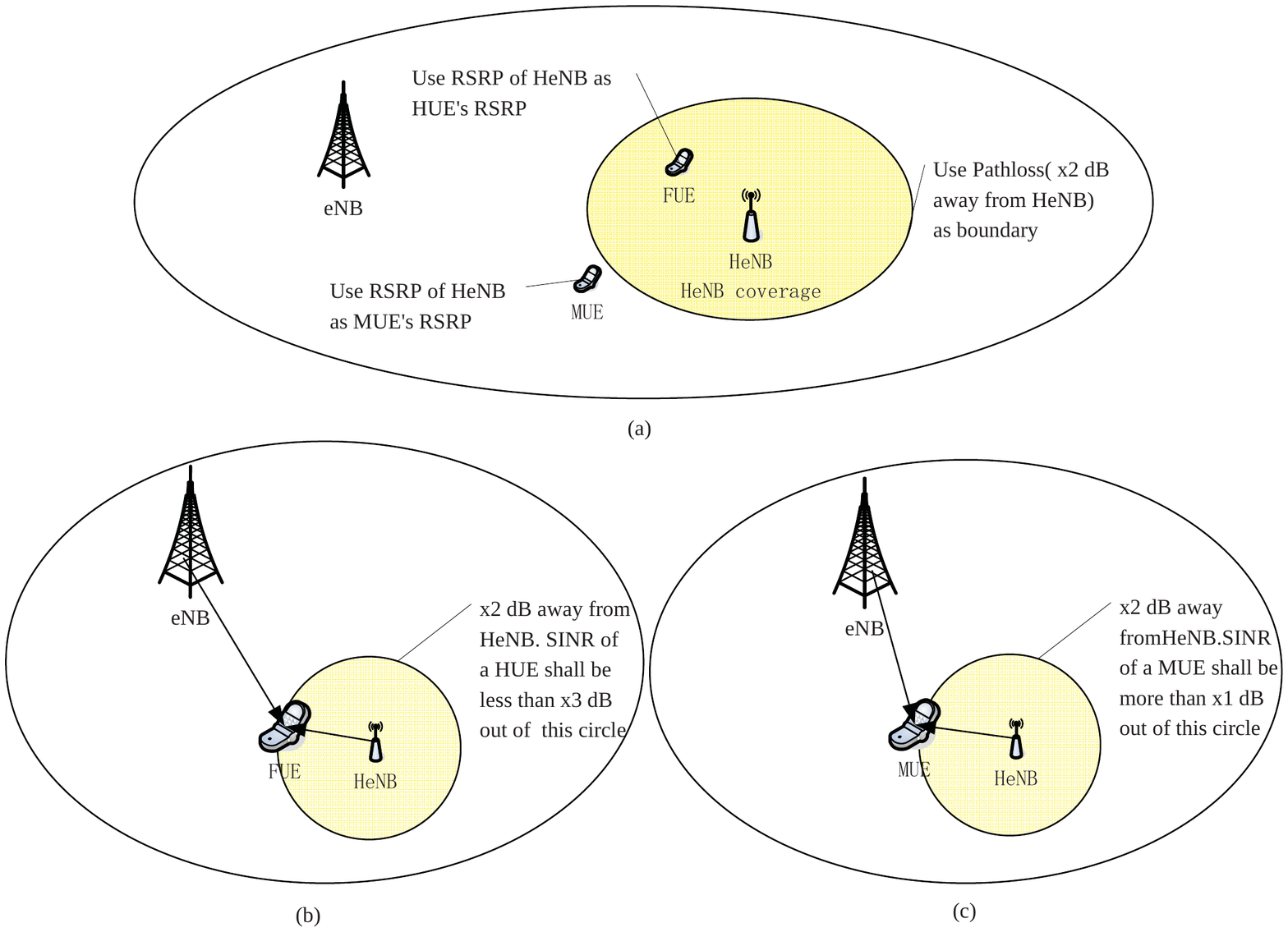}
        \caption{Self-configuration of transmitter power.}
        \label{fig:2}
\end{figure}

After power-on, HeNB can detect the wireless environment with a downlink receiver. The HeNB can get the Reference Signal Received Power (RSRP) of the macro eNB and its neighbor HeNBs from this detection. The self-configuration process will use the detection result to configure a suitable transmitter power for the HeNB with the consideration of co-channel interference mitigation. As shown in Fig. 2(a), to guarantee a self-configuration scheme to work well, we need the following assumptions 1) HeNB has a small coverage area and the boundary is defined as the place that pathloss is 2 dB away from HeNB; 2) A close-by macrocell UE (MUE) or femtocell UE (FUE) has a similar RSRP as the detection of HeNB; 3) If the SINR of a FUE received from the HeNB is more than 3 dB, it means FUE is covered by a HeNB; 4) If the SINR of a MUE received from the macro eNB is more than 1 dB, it means that MUE is covered by a macro eNB.

With the above assumptions, the basic concept of the algorithm in self-configuration process can be described as following: 1) Maintain an SINR to be less than 3 dB for a FUE located more than 2 dB away from HeNB; 2) Maintain an SINR to be more than 1 dB for a MUE located more than 2 dB away from HeNB.
The power self-configuration scheme is given in Fig. 2(b) and Fig. 2(c). Fig. 2(b) shows that the suitable transmit power of HeNB should guarantee FUE on the boundary, i.e., FUE's SINR is less than 3 dB. Thus the HeNB cannot affect the MUE. Fig. 2(c) shows the suitable transmit power of HeNB should guarantee the MUE on the boundary, i.e., MUE's SINR is more than 1 dB. The performance of the proposed algorithm is evaluated together with coverage and capacity optimization in Section IV.

\subsection{Self-Configuration of Physical Cell ID}
Basically, a physical cell ID corresponds to a unique combination of one orthogonal sequence and one pseudo-random sequence, and 504 PCI support. 504 PCIs are usually grouped into 168 unique PCI groups, each group containing three unique identities \cite{3GPP36902}. Moreover, since synchronization signals are generated by PCI, having the same PCI implies having the same synchronization signal. From that point of view, having two small cells within certain inter-cell distance will cause their respective
synchronization signals to be indistinguishable. In this situation, UEs cannot decode synchronization signals.
\begin{figure}[h!]
        \centering
        \includegraphics*[width=16cm]{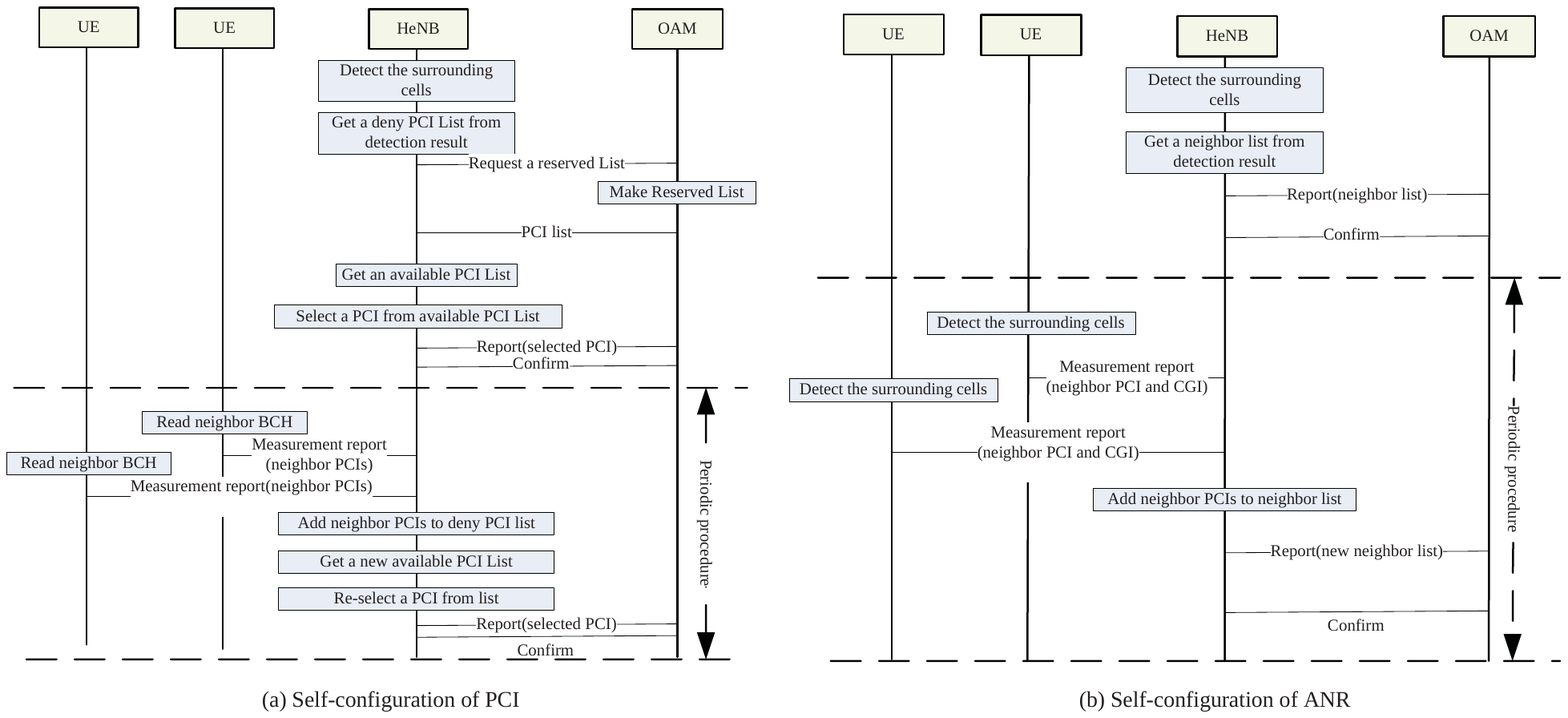}
        \caption{Self-configuration of PCI and ANR.}
        \label{fig:3}
\end{figure}

When a new small cell base station is brought into the field, a PCI needs to be selected for each of its supported cells to avoid collision with neighboring cells.  The use of one PCI by two cells would result in  hindering the identification. Traditionally, the proper PCI is derived from radio network planning and is part of the initial configuration of the node. The PCI assignment must satisfy the following conditions: i) ``collision-free", ii) the PCI is unique in the area that the cell covers, iii) ``confusion-free", iv)  a cell shall
not have a same PCI with its neighboring cells. However, it is not always possible to guarantee ``confusion-free" in a dense heterogeneous small cell deployment scenario, since there are too many overlaid HeNBs deployed under a macro eNB and the number of possible PCIs is limited to 504.

In this subsection, we present a simple but effective PCI self-configuration scheme shown in Fig. 3(a).

\begin{figure}[h!]
        \centering
        \includegraphics*[width=15cm]{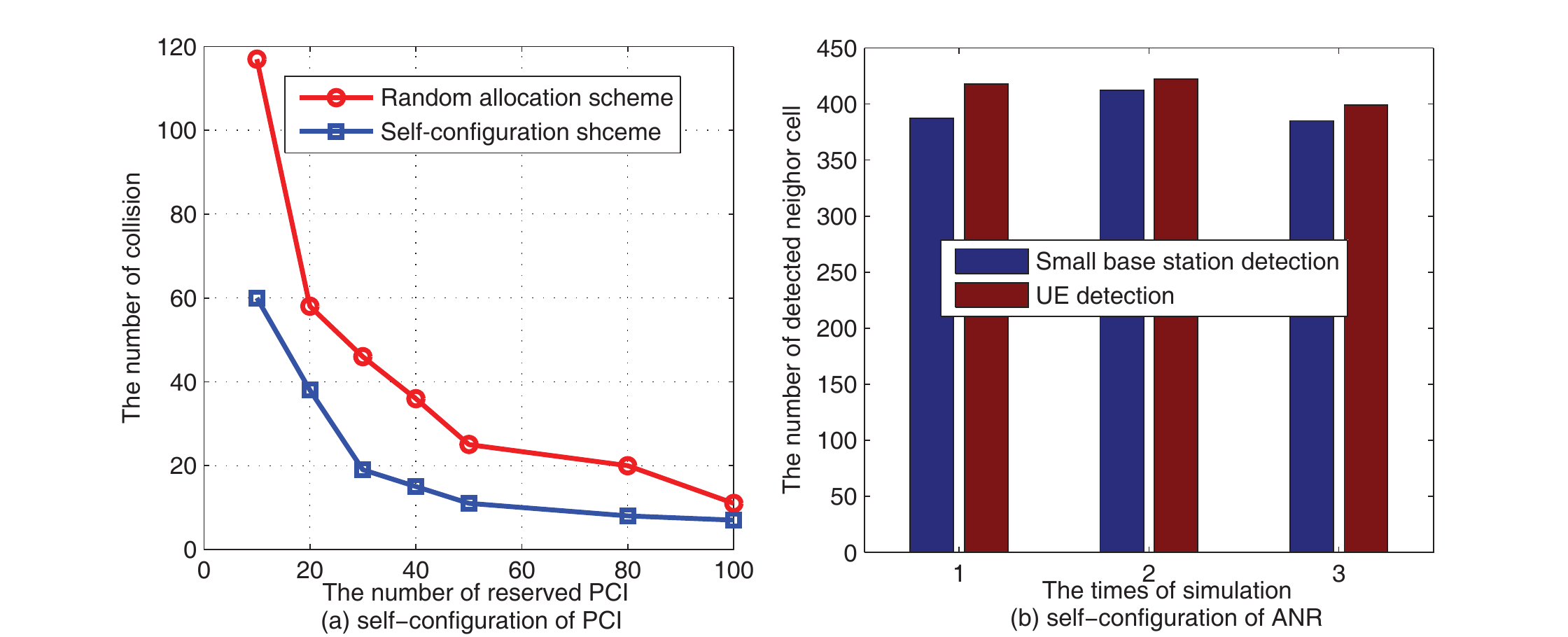}
        \caption{Simulated performance of self-configuration of PCI and ANR.}
        \label{fig:4}
\end{figure}

As can be seen from Fig. 4(a), PCI self-configuration has a lower number of collisions than the random allocation scheme in terms of the number of collisions. In other words, the proposed scheme can greatly decrease PCI confusion and collision. This is very important in disaster scenarios since small cell base stations are randomly deployed.

\subsection{Self-Configuration of ANR}

Automatic Neighbor Relation implementation is a key feature for SON enabled DRHSCNs. The purpose of the ANR function is to relieve operators from the burden of manually managing Neighbor Relations (NRs), since people may not be available in disaster scenarios \cite{3GPP36902}.
The function of ANR should be in the eNB and have power to manage the Neighbor Relation Table (NRT). In function of ANR, new neighbors are found by the Neighbor Detection Function and adds them to NRT. On the other hand, ANR can also remove the outdated NRs by using of Neighbor Removal Function.

The ANR self-configuration procedure for DRHSCN in disaster scenarios is shown in Fig. 3(b). And the performance of self-configuration of ANR is shown in Fig. 4(b). As shown in the figure, small cell base station detection based ANR can reduce the number of detected neighbor cells, which can alleviate the burden of neighbor cell selection in handover.

\section{Self-Optimization in DRHSCNs}
Handover is one of the most resource-consuming procedures so that handover needs careful design optimization by SON. From time to time, the combination of diverse user mobility patterns and complex cell boundary coverage can generate frequent unnecessary handovers that lead to RLF and unnecessary handovers. Therefore, handover related parameter optimization, such as, handover decision or access control should detect those scenarios and solve it. Since the objective of reducing unnecessary handovers  sometimes is contrary to the objective of reducing the number of handover failures, optimization algorithm design should also be able to consider the tradeoff between handover failure and unnecessary handover.

\subsection{Mobility Robustness Optimization}

In the following, we present a cost function based mobility robustness scheme \cite{EurasipMobility2013}, where the cost function is comprised of a weighted sum of the number of ping-pongs, continue handovers, late
handovers, early handovers and wrong handovers.

Ping-pong handover, continue handover, late handover, early handover and wrong handover are defined as follows \cite{IEICEMobility2012}, 1) Ping-pong handover: A handover to the serving cell from the target cell shortly after
a successful handover to the target cell;  2) Continue handover: A handover to another cell (neither the serving cell nor the target cell) shortly after a successful handover to the target cell;  3) Late handover: A RLF occurs in a serving cell before handover or during the handover procedure, and then the UE reconnects to the target cell (different from the serving cell);  4) Early handover: A RLF occurs shortly after a successful handover to the target cell, and
then the UE reconnects to the serving cell;  5) Wrong handover: A RLF occurs shortly after a successful handover to the target cell, and then the UE reconnects to another cell (neither the serving cell nor the target cell).

Also, in order to detect too late, too early, wrong handovers and call drops to construct the cost function, the following procedure is applied. In the procedure, eNB starts the timer for each UE at the moment of receiving the handover completion from each UE. During the connecting time period, if the eNB receives the RLF report from other eNBs, the eNB should stop the timer. Based on the performance metric definitions, according to UE's status after the RLF, eNB categorizes RLF as a call drop, too late handover, too early handover or wrong handover.

The optimization algorithm embedded within a single (H)eNB, collects the performance metrics and computes the
optimized parameters. The detailed optimization algorithm procedure is described in Fig. 5(a).
\begin{figure}[h]
        \centering
        \includegraphics*[width=15cm]{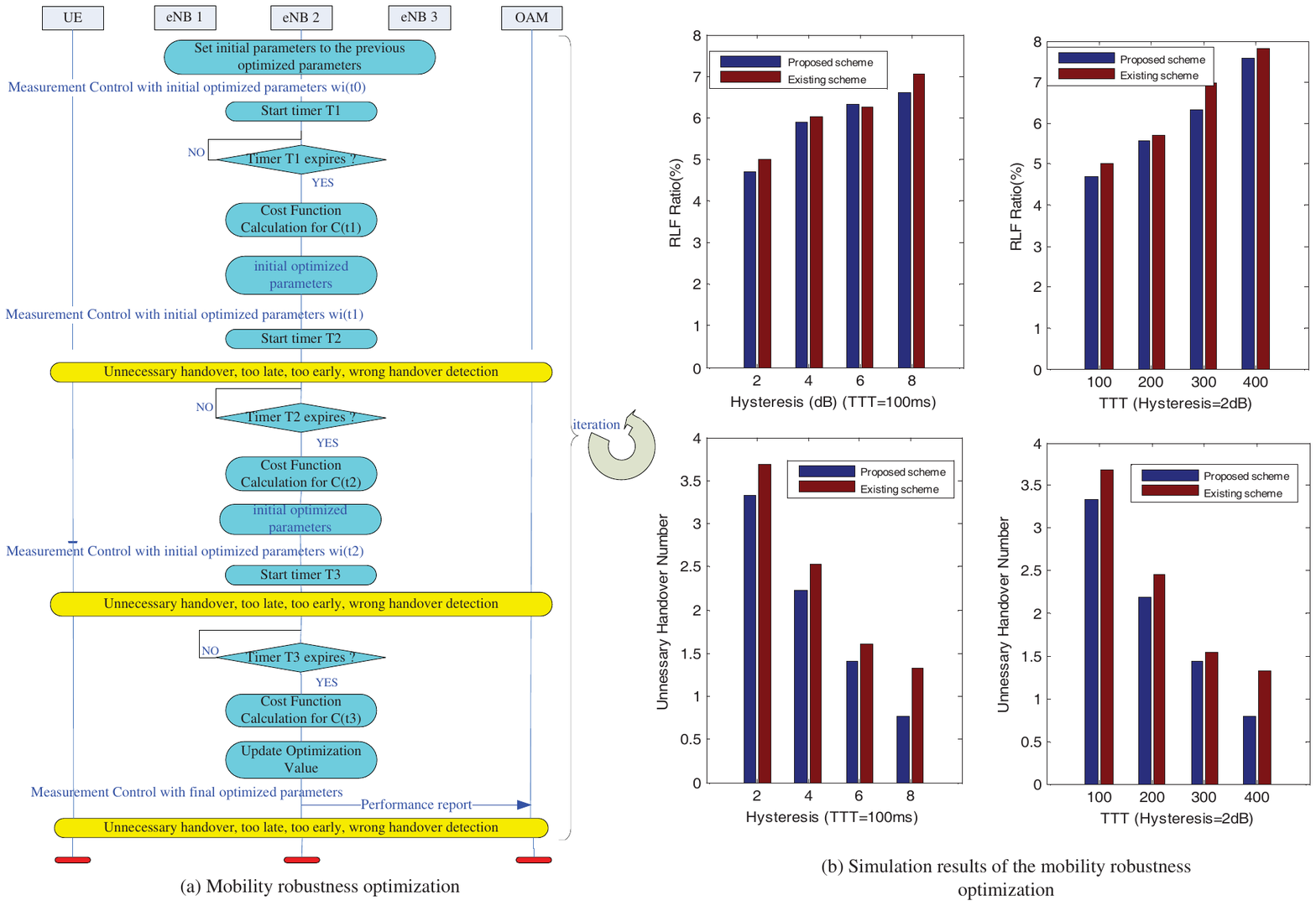}
        \caption{The MRO Optimization procedure and the simulated performance in DRHSCNs.}
        \label{fig:5}
\end{figure}

Fig. 5(b) shows a performance comparison of the proposed scheme and an existing scheme in terms of RLF radio and unnecessary handover ratio in DRHSCNs. As can be seen from the figure, the proposed cost function based mobility management scheme has a better performance than the existing scheme, which indicates that the proposed mobility robustness optimization (MRO) scheme can improve the mobility handover performance in disaster area to support the user mobility and the random deployment of small cell base stations.

\subsection{Coverage and Capacity Optimization}
Coverage and capacity optimization is an important issue for heterogeneous small cells in disaster scenario.  With the demand of small cell base station deployment in the disaster area, the interference between a macro base station and a small cell base station becomes the major obstacle for deployment, which may decrease the performance of the disaster resilient heterogenous communication networks.

In this section, we propose a boundary based coverage and capacity optimization scheme. In the self-configuration process, the boundary is assumed fixed, whereas in a real environment, the size of an apartment or a house varies.  Meanwhile, it is desirable to get the real boundaries of every HeNB coverage area. In the self-configuration process, the SON system cannot get any information except the RSRP of macro eNB and neighbor HeNB, and one cannot determine the exact boundary only from
the detection.  However, in the self-optimization process there are FUEs working in the coverage area of
HeNB, so one can use the measurement of FUE to determine the boundary.  When  a FUE enters
an apartment, its RSRP from HeNB will have a big change because of the penetration loss of
walls, so one can use the path-loss  to estimate the radius of the HeNB coverage.  The detailed procedure is described in  Fig. 6(a).
\begin{figure}[h]
        \centering
        \includegraphics*[width=15cm]{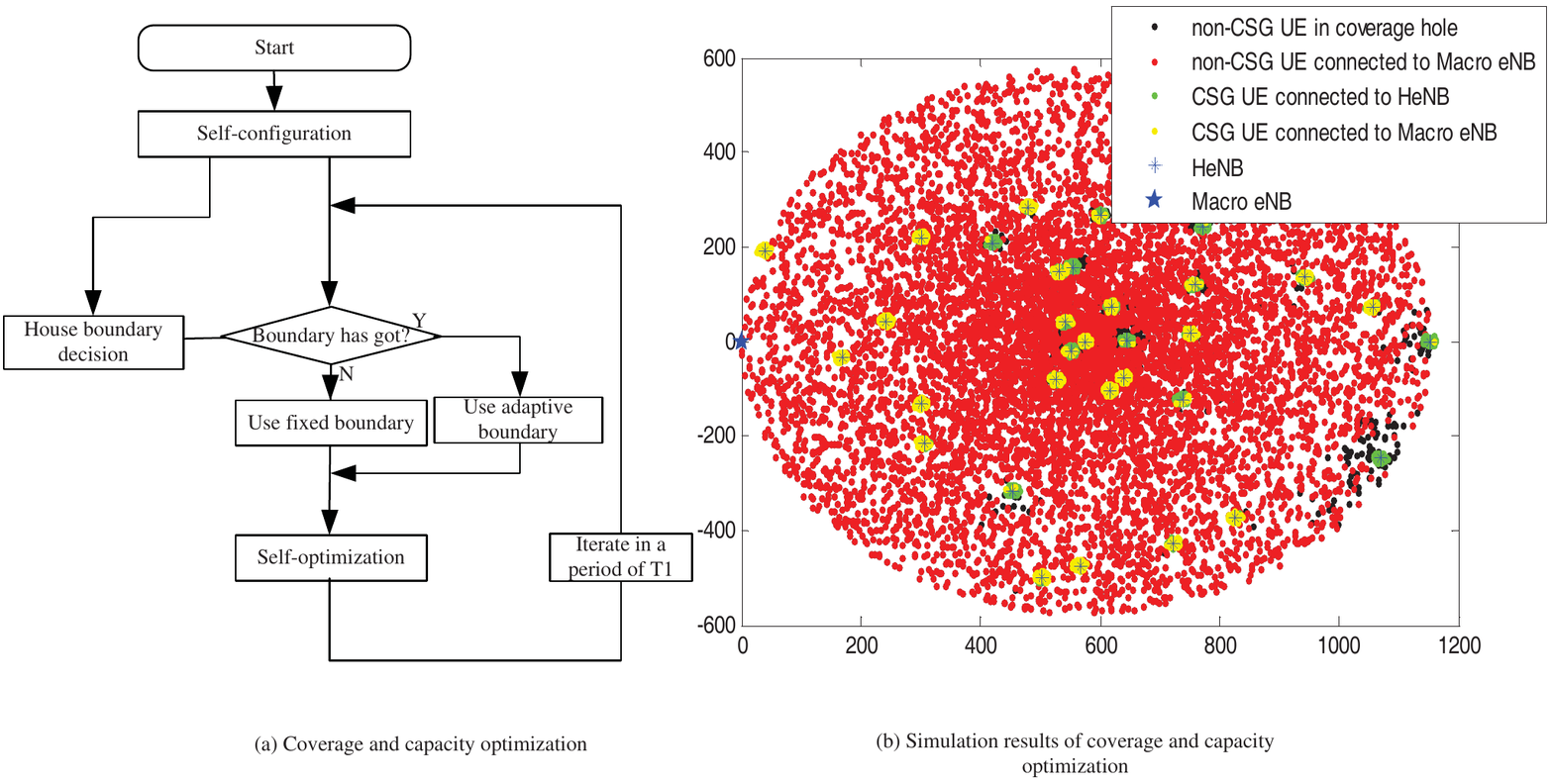}
        \caption{Coverage and capacity optimization and simulation results in DRHSCNs.}
        \label{fig:6}
\end{figure}

Fig. 6(b) shows a performance evaluation of the proposed  coverage and capacity optimization scheme (including the power self-configuration scheme proposed in Section III) in disaster scenario. As can be seen from the figure, the proposed coverage and capacity optimization scheme has a good performance in terms of macrocell coverage ratio and small cell coverage because of the self-configuration and optimization of the small cell's transmit power. Therefore, the coverage holes and capacity of wireless communication networks in disaster area can be further improved.

\section{Conclusion}
In this article, we have applied the idea of SON to enhance the disaster resilience in heterogeneous small cell networks. The applied SON techniques include self-configuration of power, PCI, and ANR, as well as self-optimization of mobility robustness and coverage and capacity. Our detailed study and performance results show that self-organization networking is an appropriate and efficient tool to handle the deployment, configuration and optimization of additional deployed small cell base stations in disaster communication networks. In the future, we will also study the self-healing technology in self-organizing disaster heterogeneous small cell networks.

\section*{Acknowledgment}
This work was supported by the National Natural Science Foundation of China (61471025, 61371079), the Fundamental Research Funds for the Central Universities (Grant No. ZY1426), and the Interdisciplinary Research Project in BUCT.

\begin{IEEEbiography}
{Haijun Zhang} (M'13) received his Ph.D. degree in School of Information and Communication Engineering (SICE), Beijing Key Laboratory of Network System Architecture and Convergence, Beijing University of Posts Telecommunications (BUPT). Currently, he is a Postdoctoral Research Fellow in Department of Electrical and Computer Engineering,  the University of British Columbia (UBC). He is also an Associate Professor in College of Information Science and Technology, Beijing University of Chemical Technology. From September 2011 to September 2012, he visited Centre for Telecommunications Research, King's College London, London, UK, as a joint PhD student and Visiting Research Associate. Dr. Zhang has published more than 50 papers and authored 2 books. He serves as editor of Wireless Networks and KSII Transactions on Internet and Information Systems. He served as Symposium Chair of the 2014 International Conference on Game Theory for Networks (GAMENETS'14) and Track Chair of 15th IEEE International Conference on Scalable Computing and Communications (ScalCom2015). He also serves as TPC members of Globecom and ICC. His current research interests include 5G, Radio Resource Management, Heterogeneous Small Cell Networks and Ultra-Dense Networks.

\end{IEEEbiography}

\begin{IEEEbiography}
{Chunxiao Jiang} (S'09, M'13) received the B.S.  degree (with highest honors) in information engineering from Beihang University (formerly Beijing University of Aeronautics and Astronautics), Beijing, China, in 2008 and the Ph.D. degree (with highest honors) from Tsinghua University (THU), Beijing, in 2013. During 2011¨C2013, he visited the Signals and Information Group, Department of Electrical and Computer Engineering, University of Maryland, College Park,MD, USA, with Prof. K. J. Ray Liu. He is currently a Postdoctoral Fellow at the Department of Electronic Engineering, THU, with Prof. Y. Ren. His research interests include the applications of game theory and queuing theory in wireless communication and networking and social networks. Dr. Jiang was a recipient of the Best Paper Award at IEEE GLOBECOM in 2013, the Beijing Distinguished Graduated Student Award, Chinese National Fellowship, and Tsinghua Outstanding Distinguished Doctoral Dissertation in 2013.

\end{IEEEbiography}

\begin{IEEEbiography}
{Rose Qingyang Hu} (S'95, M'98, SM'06) received a B.S. degree in Electrical Engineering from University of Science and Technology of China, a M.S. degree in Mechanical Engineering from Polytechnic Institute of New York University, and a Ph.D. degree in Electrical Engineering from the University of Kansas. From January 2002 to June 2004 she was an assistant professor with the Department of Electrical and Computer Engineering at Mississippi State University. She also has more than 10 years of R\&D experience with Nortel, Blackberry and Intel as a technical manager, a senior research scientist, and a senior wireless system architect. She was Nortel 4G system level simulation prime and led Notel 4G standards and technology performance evaluation. Currently she is an associate professor with the Department of Electrical and Computer Engineering at Utah State University. Her current research interests include next-generation wireless communications, wireless network design and optimization, green radios, multimedia QoS\/QoE, communication and information security, wireless system modeling and performance analysis. She has published extensively and holds numerous patents in her research areas. She is currently serving on the editorial boards for IEEE Wireless Communications Magazine, IEEE Internet of Things Journal, IEEE Communications Surveys and Tutorials. She has  been a 6-time guest editor for IEEE Communications Magazine, IEEE Wireless Communications Magazine, and IEEE Network Magazine.  Prof. Hu received IEEE Globecom 2012 Best Paper Award.  She is IEEE ComSoc Distinguished Lecturer 2015-2016, a senior member of IEEE,  and a member of Phi Kappa Phi and Epsilon Pi Epsilon Honor Societies.
\end{IEEEbiography}

\begin{IEEEbiography}
{Yi Qian} (M'95, SM'07) is an associate professor in the Department of Electrical and Computer Engineering, University of Nebraska-Lincoln (UNL). Prior to joining UNL, he worked in the telecommunications industry, academia, and the government. Some of his previous professional positions include serving as a senior member of scientific staff and a technical advisor at Nortel Networks, a senior systems engineer and a technical advisor at several start-up companies, an assistant professor at University of Puerto Rico at Mayaguez, and a senior researcher at National Institute of Standards and Technology. His research interests include information assurance and network security, network design, network modeling, simulation and performance analysis for next generation wireless networks, wireless ad-hoc and sensor networks, vehicular networks, smart grid communication networks, broadband satellite networks, optical networks, high-speed networks and the Internet. He has a successful track record to lead research teams and to publish research results in leading scientific journals and conferences. Several of his recent journal articles on wireless network design and wireless network security are among the most accessed papers in the IEEE Digital Library.
\end{IEEEbiography}

\end{document}